\documentstyle[epsfig]{mn2e}

\title[High mass accretion in GX~339$-$4]
{Modelling the high mass accretion rate spectra of \\
GX~339$-$4: Black hole spin from reflection?}

\author[Kolehmainen, Done \& D{\'{\i}}az Trigo]
{Mari Kolehmainen, $^1\thanks{E-mail:m.j.kolehmainen@durham.ac.uk}$ Chris Done $^1$ and Mar{\'{\i}}a D{\'{\i}}az Trigo $^{2}$\\
$^1$Department of Physics, University of Durham, South Road, Durham
DH1 3LE,
UK\\
$^2$ European Southern Observatory, ALMA Regional Centre, Karl-Schwarzschild-Str. 2, D-85748 Garching, Germany \\
}

\pagerange{\pageref{firstpage}--\pageref{lastpage}} \pubyear{2010}

\begin{document}

\topmargin = -0.5cm

\maketitle

\label{firstpage}

\begin{abstract}

We extract all the \textit{XMM-Newton} EPIC pn burst mode spectra of GX~339$-$4,
together with simultaneous/contemporaneous \textit{RXTE} data.
These include three disc dominated and two
soft intermediate spectra, and the combination of broad
bandpass/moderate spectral resolution gives some of the best data on these
bright soft states in black hole binaries.
The disc dominated spectra span a factor three in luminosity, and
all show that the disc emission is broader than the simplest multicolour
disc model. This is consistent with
the expected relativistic smearing and changing colour temperature
correction produced by atomic features in the newest disc models.
However, these models do not match the data at the 5 per cent level as the
predicted atomic features are not present in the data, perhaps indicating
that irradiation is important even when the high energy tail
is weak. Whatever the reason, this means that the data have smaller errors
than the best physical disc models, forcing use of more phenomenological
models for the disc emission. We use
these for the soft intermediate state data, where previous analysis using
a simple disc continuum found an extremely broad residual, identified as the
red wing of the iron line from reflection around a highly spinning black
hole. However, the iron line energy is close to where the disc and tail
have equal fluxes, so using a broader disc continuum changes the residual
'iron line' profile dramatically. With a broader disc continuum model, the
inferred line is formed outside of 30~${\rm{R_g}}$, so cannot constrain black hole
spin. We caution that a robust determination of black hole spin from the
iron line profile is very difficult
where the disc makes a significant contribution at the iron line energy i.e.
in most bright black hole states.

\end{abstract}

\section{Introduction}

Black hole spin is currently a subject of intense debate as it is very
difficult to measure. Unlike mass, spin only leaves an imprint on the
spacetime very close to the event horizon. Nonetheless, there are now
two ways to study this. The first uses the temperature and luminosity
of the accretion disc formed by material falling into the black hole.
These parameters are determined by the combination of the rate at
which material is swept through the disc, and how far down the disc
can extend close to the black hole. A spinning black hole drags
spacetime around with it, so that the accretion disc can extend closer
in. This gives higher temperature and luminosity for a given accretion
rate, or equivalently in terms of observables, a higher disc
temperature for a given luminosity (e.g. Done, Gierli\'nski \& Kubota, 2007). 

The second method uses the profile of an iron line produced by
fluorescence where the X-rays illuminate the accretion disc. The
closer the disc extends down to the black hole, the faster it orbits
around, and the stronger the effects of special and general
relativity. The large velocity and strong gravity sculpt the line
profile, broadening it from a narrow atomic transition in a way which
can now be observed (e.g the review by Fabian et al. 2000). 

It is obviously important to compare results from these two methods.
However, the current level of agreement is not very encouraging.  One
of the most clearcut cases is GX~339$-$4, where there are claims of
extremely high spin from a very broad iron line in 3 datasets, but
where the disc continuum fits strongly prefer lower spin. However, two
of the three broad iron line detections have been challenged as being
due to instrumental pile-up (Miller et al. 2006 vs. Done \& Diaz Trigo
2010 and Miller et al. 2008 vs Yamada et al. 2009). Pile-up occurs where
the source is so bright that there is a high probability of two
photons hitting either a single pixel within the readout time of that
pixel, so that the sum of the energies is assigned to a single photon
(energy pile-up) or of two photons hitting adjacent pixels, and being
treated as a single photon split between two pixels (double event) with
summed energy (pattern pile-up). Both processes distort the spectrum in
a way which is not well understood, making detailed spectral fitting
(as required for the iron line profile) difficult. 

However, the remaining observation of GX~339$-$4 in which the very broad
line is seen was taken in the burst mode of \textit{XMM-Newton}, which is
especially designed to handle the highest count rates without
pile-up. It does this by reducing the readout time, but at the cost of
a dramatic reduction in the detector live time. Only 3\% of the
available photons are collected, but the spectrum is free from
the uncertainties associated with pile-up. However, while this fast
timing burst data mode has been available for the \textit{XMM-Newton}
EPIC pn detector since the launch, it has only recently been
calibrated well enough to give reliable spectra (see e.g. Guainazzi et al. 2010:
XMM-SOC-CAL-TN-0083\footnote{http://xmm.vilspa.esa.es/external/xmm\_sw\_cal/ \\
calib/documentation.shtml}).

Here we re-examine all the \textit{XMM-Newton} burst mode spectra 
of GX~339$-$4 to
assess the level of agreement between the spin derived from disc
continuum and iron line profile. We find that the black hole spin
derived from disc fitting with CCD data is similar to that derived
from disc fitting of the higher energy \textit{RXTE} data. However, the data
also show that the best current disc contiuum models give 5--10\%
residuals, as they predict (smeared) atomic absorption features in the
disc photosphere which are not present in the data.  This is unlikely to be due to
any remaining calibration issues in this mode of \textit{XMM-Newton}, or to
interstellar absorption, as similar residuals are seen in \textit{Suzaku} data
from LMC X-3, which has a very low galactic column (Kubota et al.
2010). Instead it seems more likely that illumination (either self
illumination of the inner disc by its own emission or by the hard
tail) drives the photosphere towards isothermality.  Whatever the
underlying cause, the disc dominated spectra are clearly broader than
expected from an emissivity weighted sum of blackbody spectra, such as the {\sc
diskbb} model (Mitsuda et al. 1984). 

The shape of the disc spectrum is especially important when it comes
to disentangling the iron line profile from the continuum. Fitting a
narrow {\sc diskbb} shape to the accretion disc emission forces a
broad residual into the data. Similarly, the high energy continuum is
not a simple power law (Kubota et al. 2001; Zycki, Done \& Smith 2001, 
Gierli\'nski et al. 1999), so fitting such models again forces a broad residual into the
spectrum.  The iron line shape derived from our fits including a
broader disc spectrum and a Comptonised continuum is not extremely
broad. It is easily consistent with the lower spin derived from the
disc spectral fitting. We caution that the continuum is complex in the
soft states, especially the brighter soft states where the disc
extends to 6--7~keV, and that how the continuum components
are modelled makes a
difference to the derived profile of the iron line.

\section{Observations and data reduction}

\begin{figure}
\begin{center}
\leavevmode\epsfxsize=8cm\epsfbox{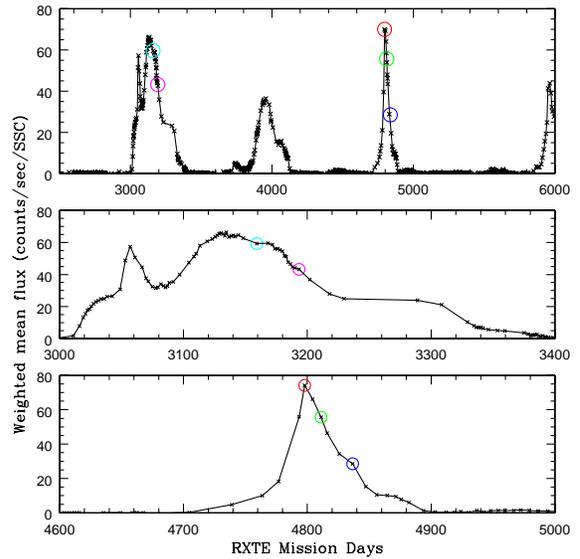}
\end{center}
\caption{The long term \textit{RXTE} ASM light curve of GX~339$-$4
(top panel) with the 2002/2003 and 2007 outbursts scaled in (middle and
bottom panels, respectively). Times corresponding to the \textit{XMM-Newton}
burst mode observations are shown by the coloured circles.
}
\label{lcurve}
\end{figure}

GX~339$-$4 reaches a flux of $\sim 10^{-8}$ ergs cm$^{-2}$ s$^{-1}$,  
i.e. $\sim$0.5 Crab (e.g. Dunn et al. 2008). This corresponds to count rates
in \textit{XMM-Newton} of $\sim 5000$ c/s, well above the nominal 
pile-up limit of the timing mode of 800 c/s. Thus the only option 
for robust spectral fitting is the burst mode. Table ~\ref{obs} gives details
of all burst mode observations of GX~339$-$4, 
along with details of the \textit{RXTE} dataset closest in time to each
observation. Figure ~\ref{lcurve} shows where these spectra fall on the
long term \textit{RXTE} PCA lightcurve of this source, with 2 datasets in the
2002/2003 and 3 in the 2007 outbursts.

\begin{table}
\begin{tabular}{lc|l|l|l|l}
 \hline
   \small & {Obsid} & \small{Date} & \small{Exp(s)} &\small{State}\\ 
\hline
 & & \small {EPIC pn} \\
  \hline
1 & 0093562701  & 2002-08-24 10:30:16 & 60640 & Disc dom. \\
2 & 0156760101  & 2002-09-29 08:56:46 & 75601 & SIMS	\\
3 & 0410581201  & 2007-02-19 00:03:25 & 15048 & Disc dom.\\
4 & 0410581301  & 2007-03-05 11:15:56 & ~3200 & SIMS	\\
5 & 0410581701  & 2007-03-30 15:01:07 & ~8658 & Disc dom.\\
\hline
 & & \small {\textit{RXTE} PCA} \\
\hline
1 & 70130-01-01-00  & 2002-08-24 10:43:07.1 & 3564 & Disc dom.\\
2 & 70130-01-02-00  & 2002-09-29 12:10:17.8 & 9792 & SIMS\\
3 & 92085-01-01-00  & 2007-02-19 17:41:17.7 & 3504 & Disc dom.\\
4 & 92085-01-03-03  & 2007-03-05 13:16:48.6 & 3200 & SIMS\\
5 & 92085-02-03-00  & 2007-03-30 00:57:49   & 3500 & Disc dom.\\
\hline
\end{tabular}
\caption{Details of the observations. The quoted exposure times are the exposures used in this analysis.}
\label{obs}
\end{table}

\begin{figure}
\begin{center}
\leavevmode\epsfxsize=8cm\epsfbox{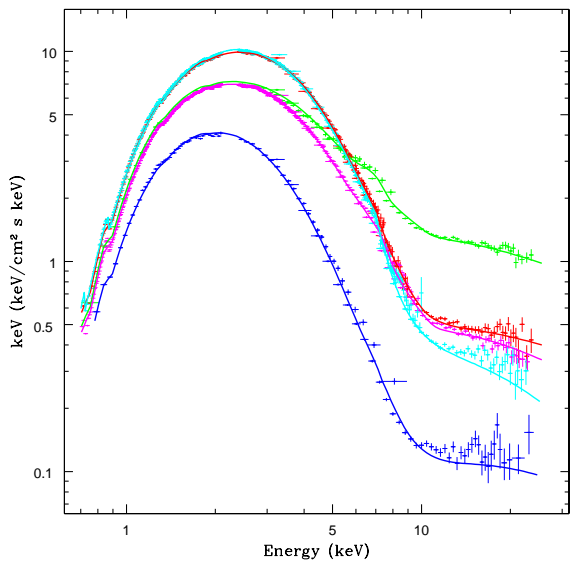}
\end{center}
\caption{The high mass accretion rate spectra of GX~339$-$4. 
Colour convention same as in Figure ~\ref{lcurve}: Obs 1. in cyan, Obs 2. in magenta, Obs 3. in red, Obs 4. in green and Obs 5. in blue.}
\label{all_spectra}
\end{figure}

The pn burst mode data were reduced using the latest version of the
\textit{XMM-Newton} Science Analysis System (SAS) v10.0.  While the
count rates of 1000--5000 c/s ($\sim 0.1-0.5$ Crab) are well below the
nominal pile-up limit of 60,000 c/s of this mode, an analysis of the
Crab nebula showed that sources of this intensity are somewhat
affected by pile-up at 140$\leq$RAWY$\leq$180 due to the special
readout (Kirsch et al. 2006). We follow their recommendation for
robust spectral determination, i.e. extract events in RAWY
[1:160]. Slightly different positioning on the chip for the different
observations means that we use either RAWX [31:41] or RAWX [31:42]. We
only use integer column numbers as the SAS tasks for generating
response and ancillary files ({\sc rmfgen} and {\sc arfgen}) silently
truncate any fractional column number. We use only single and double
events (PATTERN $\leq$ 4) and ignore band pixels with \#XMMEA\_EP and
FLAG==0. We correct for rate-dependent charge transfer inefficiency
(CTI), using {\sc epfast}, which is included in the latest version of
SAS. CTI occurs where electrons are caught in charge traps rather than
being read out. This causes small shifts to the energy gain, so is
most noticable where the effective area of the instrument changes
i.e. at the Si and Au features. The level of residuals at these
energies is a measure of success of the CTI correction. Our residuals around the Au edge are at the 2-5\% level expected for the current calibration, but this is larger than the 1\% systematic error applied to the rest of the spectrum,
so we exclude this region from the fit.

Finally the spectra were rebinned using the SAS task {\sc specgroup}. The number of bins per instrumental energy resolution was set to 3, as recommended for the EPIC pn, to make sure that all the channels are indeed independent for the $\chi^2$ calculations. The high signal-to-noise at low energies and the high oversampling of the \textit{XMM-Newton} response will otherwise lead to the spectral fits to be weighted more towards these low energies than the high, and hence artificially low $\chi^2$ values. Each bin was also set to have minimum of 25 counts. Background was not extracted as all regions on the chip are contaminated by the source (Done \& Diaz Trigo 2010, see Appendix~\ref{bkg}).

For each \textit{XMM-Newton} observation we also extract the RXTE data nearest in
time. These all have significantly shorter exposure than the \textit{XMM-Newton}
datasets. Observations 1, 2 and 4 are all within the corresponding
\textit{XMM-Newton} dataset, but 3 and 5 have no overlap in time. However, disc
dominated states are known to have rather slow variability, so variability
is probably only a potential issue for the observation with
the most dominant hard tail (Obs 4). Hence we
time filter this \textit{XMM-Newton} dataset to the \textit{RXTE} observation to get truely
simultaneous data. We allow there to be a
free normalisation between the PCA and \textit{XMM-Newton} data, but tie all
the spectral parameters across the two datasets.

\begin{table*}
\begin{center}
\begin{tabular}{l|l|l|lc}
\hline
\textbf{Table 2.} & {\sc the models} & &  \\
\hline
& Model 1a & {\sc tbabs*smedge*(diskbb+thcomp+gaussian)} & &  \\
& Model 1b & {\sc tbabs*smedge*(bhspec+thcomp+gaussian)}  & & \\
& Model 2  & {\sc tbabs*(bhspec+thcomp+kdblur*refxion*(thcomp))}  \\
& Model 3  & {\sc tbabs*(simpl(bhspec)+kdblur*refxion*simpl(bhspec))}    \\
& Model 4  & {\sc tbabs*(simpl(diskbb+comptt)+kdblur*refxion*simpl(diskbb+comptt))}  \\
 \hline
\end{tabular}
\caption{}
\end{center}
\label{models}
\end{table*}

\section{Spectral analysis overview}

We previously analysed all the disc dominated \textit{RXTE} PCA data from GX~339$-$4 
using a multicolour disc and its thermal Comptonisation,
together with a Gaussian line and smeared edge to approximately model
the expected reflection features (({\small {\sc tbabs*smedge*(diskbb+thcomp+gauss)}}):
Kolehmainen \& Done 2010, Model 1a in Table~\ref{models}).  
We now test this same model on our composite \textit{RXTE/XMM-Newton}
datasets which extends the low energy bandpass to 0.7~keV. We could
include the simultaneous \textit{RXTE} HEXTE data in the fit, but the much
lower signal to noise at these high energies mean that these points
are given very little weight in the spectral fitting. Instead, we
include these data after the fit, and ratio the observed flux to the
model prediction in the 25--100~keV bandpass to assess how well the
model extrapolates to higher energies. 

Figure~\ref{all_spectra} shows the derived spectra for the joint pn-PCA datasets.
Clearly all of these have a large disc component, but formally only
observations 1, 3 and 5 (cyan, red and blue) are disc dominated, 
while observations 2 and 4 (magenta and green)
are soft intermediate states. Observations 1 and 3 have the most 
dominant disc, while observation 4 has the strongest tail. 

\begin{figure}
\begin{center}
\leavevmode\epsfxsize=8cm\epsfbox{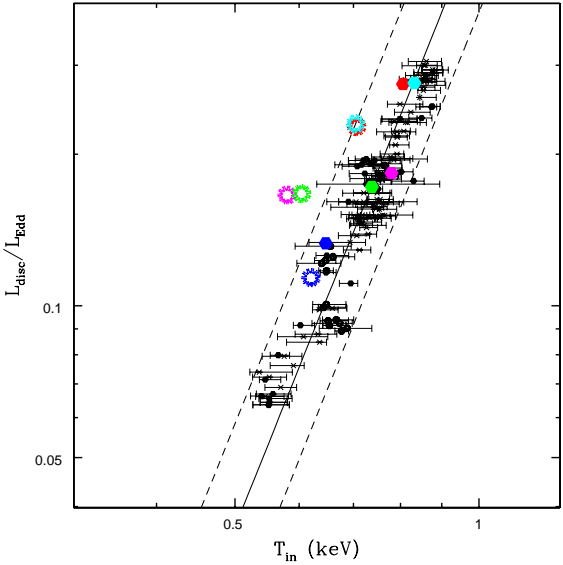}
\end{center}
\caption{The L--T$^4$ diagram of all previous disc dominated data from
the \textit{RXTE} PCA sample of 
Kolehmainen \& Done 2010 (black points) fitted with the simple Model
1a (see Table~\ref{models}). The L--T$^4$ points fitted with the same model from the
PCA data corresponding to each burst mode observation are shown as
the coloured solid symbols (same colouring convention as in Fig~\ref{lcurve}), while the
L--T$^4$ points from joint pn-PCA fits are shown as circles. These show
a lower colour-temperature correction, as expected, but lie progresively
further from the line as the strength of the hard tail increases.
}
\label{l-t}
\end{figure}

Figure~\ref{l-t} shows the disc luminosity-temperature plot from Kolehmainen
\& Done (2010) (black points) assuming a distance, mass
and inclination of 8~kpc, 10~M$_\odot$ and 60$^\circ$, respectively. 
The crosses show the parameters for the PCA data alone, derived from 
Model 1a with the absorption fixed at $6\times 10^{21}$~cm$^{-2}$
(as in Kolehmainen \& Done 2010). All these observations lie on the 
same $L\propto T^4$ relation as the disc dominated data from Kolehmainen
\& Done (2010).  The solid symbols show how this changes when the pn
data are included. Including the lower energy data
shifts the best-fit disc parameters to lower temperature/lower
luminosity, as predicted from simulations of changing bandpass with
more sophisticated disc models (Done \& Davis 2008), but
the effect is rather small. The disc dominated spectra lie closer to
the previous luminosity-temperature relation, while the soft
intermediate states are increasingly shifted, so that they have lower
temperature than expected for their luminosity (equivalent to a larger
disc radius). However, the interpretation of this is complex due to
the significant flux carried in the Comptonised tail, and correcting
the disc luminosity for these scattered photons can shift these points
back towards the constant radius disc luminosity-temperature relation
(Kubota \& Done 2004; Done \& Kubota 2006; Steiner et al. 2010)

\section{The disc dominated spectra}
\label{discy}

\subsection{The brightest disc dominated state: Obs. 3}
\label{obs3}

As the uncertainties in reconstructing the intrinsic disc properties
increase when the tail gets stronger, the disc dominated spectra
provide the clearest view of the underlying disc physics. We start our
analysis with the bright disc dominated state at the peak of the 2007 outburst (Obs 3).

The simplified model used in the previous section (Model 1a in Table~\ref{models}) 
gives acceptable results for the limited resolution of the PCA
detectors over the 3--25~keV bandpass. However, it gives a very poor
fit to the joint pn-PCA data ($\chi^2_\nu=745/185$). This is to be 
expected as the true disc is not as
simple as a multicolour blackbody. However, replacing {\small {\sc
diskbb}} with a better disc model {\small {\sc bhspec}} (Davis et
al. 2005, Model 1b; assuming M=10~M$_{\odot}$, D=8~kpc and an
inclination of $60^{\circ}$) makes the fit only moderately better
($\chi^2_\nu=530/185$). 
Thus it seems more likely that the issue is with the very approximate
description of the reflected continuum. We have combined the 
{\sc refxion} tabulated reflection models of Ross \& Fabian (2005) 
with the older {\sc pexriv} models to make a 
convolution model which can be used with any input continuum
(rather than the hardwired power law with exponential cutoff of the
{\sc refxion} models), and with normalisation given in terms of
inclination and solid angle (as for {\sc pexriv}), but with the much
more accurate treatment of the ionised reflection of {\sc refxion}
(see also an older version of this using the Ballantyne et al. 2001
tabulated reflection models in Done \& Gierlinski 2006). We 
convolve this reflected emission with the relativistic smearing
Greens functions of Laor (1991), so the total model is 
{\small {\sc tbabs*(bhspec+thcomp+kdblur2*refxion*(thcomp))}} (Model 2).
This results in another slight improvement, with
$\chi^2=505/186$, but the model clearly underpredicts the
higher energy HEXTE data, with model 25--100~keV count rate of 2.5,
compared to the 4.8$\pm$0.3 observed.

One issue with the model above is that the Compton scattering of seed
photons from the disc by a corona should also remove disc photons from
our line of sight (see the discussion of different geometries in
Kubota \& Done 2004). The Comptonisation model used here, {\sc thcomp},
does not couple the disc and tail together, so instead we replace it with 
the convolution model {\small {\sc simpl}}
(Steiner et al. 2009), which removes as many photons from the disc as
are scattered up into the tail. This model has the additional advantages
that it takes the seed photon shape from the model, rather than
assuming blackbody or {\sc diskbb} shape as for the {\sc thcomp}
model. It also assumes a power law tail, better suited to modelling
the probably non-thermal emission seen in the high/soft states than
thermal Compton scattering (Gierli\'nski et al. 1999).

However, the {\sc simpl} model, as released, does not allow the model to
be used to calculate the tail separately. This is required for the
reflection modelling, as only the coronal emission should be
reflected. Hence we re-coded the {\sc simpl} model to allow it to do
this. However, this model (Model 3 in Table~\ref{models}, with parameters
tabulated in Table~\ref{model3}) gives only a slightly better $\chi^2=426/186$ than
Model 2, with very similar parameters, including the mismatch between
the extrapolated model and the observed HEXTE flux. Thus the issue is
{\em not} with the description of the Comptonised tail (thermal vs.
non-thermal) or with the disc normalisation needing correcting for
Compton scattering. 

\begin{figure}
\begin{center}
\leavevmode\epsfxsize=8cm\epsfbox{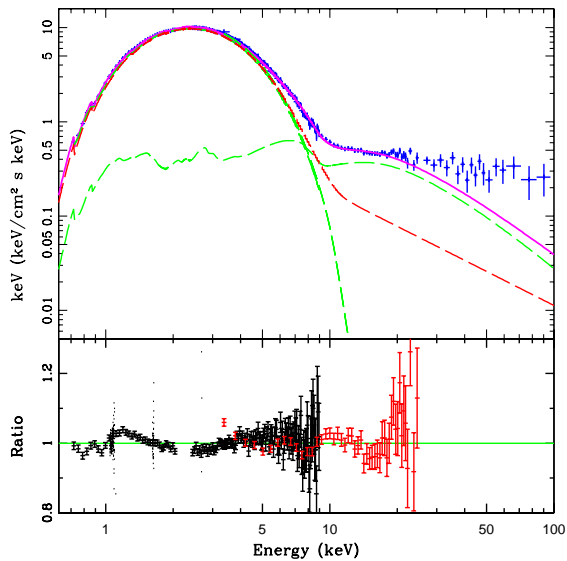}
\end{center}
\caption{The bright disc dominated spectrum (Obs. 3) modelled with Model 3. The top panel shows the data$+$model, along with the model components (disc+reflection). The model residuals are plotted in the bottom panel.} 
\label{1201_bhspec}
\end{figure}

\begin{figure}
\begin{center}
\leavevmode\epsfxsize=8cm\epsfbox{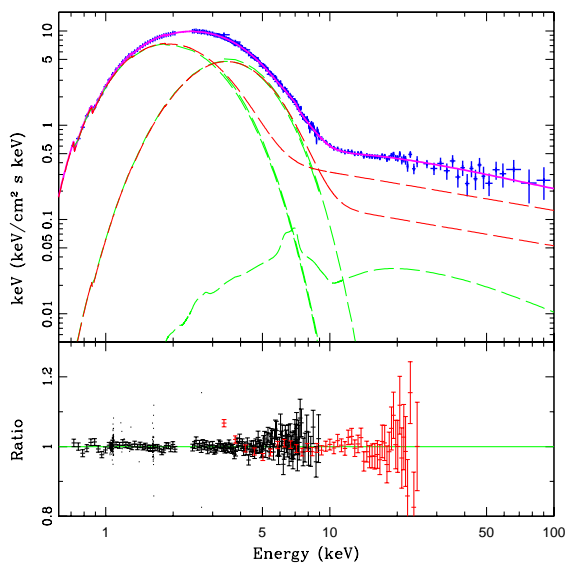}
\end{center}
\caption{Same as Fig ~\ref{1201_bhspec}, this time modelled with Model 4. } 
\label{1201_conv}
\end{figure} 

Figure~\ref{1201_bhspec} shows the deconvolved ${\nu}F_{\nu}$ 
spectrum. Plainly the fit
does actually describe the data quite well below 20~keV, but the
Compton tail has a much steeper index than expected,
at $\Gamma=3.19^{+0.09}_{-0.05}$ compared to the
typical $\Gamma\sim 2.2$ seen in the high/soft state. 
To get through the 10--20~keV PCA data then requires a
very large reflection fraction $\Omega/2\pi=6.9^{+0.4}_{-0.6}$
to flatten the spectrum, but reflection rolls over above 30--50~keV so
the extrapolation falls well below the level of the HEXTE flux. 

Thus the key issue is the derived steep Comptonisation index. One
possible reason for this is that the extremely high signal-to-noise in
the pn spectra are driving the fit, so that small residuals from the
disc model are compensated for by a steep Comptonisation tail. This
could indicate that the disc is broader than the {\sc bhspec} model,
so we replace this by a phenemological disc spectrum made from {\sc
diskbb}+{\sc comptt} (Model 4 in Table~\ref{models}). This gives a much better
fit to the data $\chi^2_\nu=210/186$, as seen by the lower level of
residuals below 1.5~keV (Figure~\ref{1201_conv}). There are still two features below
1~keV, possibly Ne III and Ne II lines from the interstellar medium
(Miller et al. 2004b), but no big residuals at iron. 

At first sight it seems likely that the better fit with the
phenomenological disc spectrum could simply be compensating for
remaining calibration uncertainties in the instrument response of this
mode. Figure~\ref{disc1201} shows a comparison of
the phenomenological {\sc diskbb+comptt} continuum with the {\sc bhspec}
continuum. Clearly the biggest difference is that {\sc bhspec}
predicts absorption features from
ionised oxygen and iron L in the disc photosphere from 0.7--1.5~keV. These
should be smeared by  relativistic effects, resulting in the broad dip
predicted by {\sc bhspec} rather than
sharp edges.The phenomenological model does not have these features, but
is as broad as the {\sc bhspec} continuum. This pattern of residuals is
very similar to that obtained from a recent \textit{Suzaku} spectrum
of LMC X-3 in a disc dominated state (Kubota et al. 2010).
These are seen even more unambiguously in LMC X-3 due to its much
smaller galactic column density along the line of sight
($0.038\times 10^{22}$ cm$^{-2}$ compared to
$\sim 0.52\times 10^{22}$ cm$^{-2}$ for GX~339$-$4).
The different instrument (\textit{Suzaku}) also
means that the issue is not likely to be simply the \textit{XMM-Newton} burst mode
calibration.

However, the shape of the derived disc continuum is
difficult to explain. It is as broad as the total continuum predicted by
the {\sc bhspec} models, but these models produce the broad continuum
partly by the changing
colour-temperature correction associated with the atomic features
yet the atomic features are not observed.
Relativistic smearing of a single colour-tempearture corrected disc
spectrum (as in the {\sc kerrbb} disc model: Li et al. 2005)
is not broad enough to explain the data (green line in Fig~\ref{disc1201}).
For comparison, we also show the best fit {\sc diskbb} model, which is
plainly far too narrow (Fig~\ref{disc1201}).

Thus it seems most likely that even the best current disc models do not describe
the observed disc spectra at the 5-10\% level, so we also use the best fit
phenomenological description (Model 4) in all the following analysis.

\subsection{The other disc dominated states: Obs. 1 \& 5}

We fit the two other disc dominated states with Models 3 and 4
described above, with best-fit parameters given in Tables~\ref{model3} and \ref{model4},
respectively. We see the same pattern as before,
namely that the best current disc models are not quite the
right shape to match with the high signal-to-noise EPIC pn data, so they
drag the Comptonisation index to higher values to compensate, which in
turn requires more reflection to flatten the spectrum in the
10--20~keV bandpass so that it can 
match the observed PCA data. This then fails to
fit the HEXTE points. By contrast, the phenomenological {\sc
diskbb+comptt} disc continuum fits the data much better, gives
more reasonable values for the Comptonisation index and amount of
reflection, and extrapolate to the HEXTE band.

\begin{figure}
\begin{center}
\leavevmode\epsfxsize=8cm\epsfbox{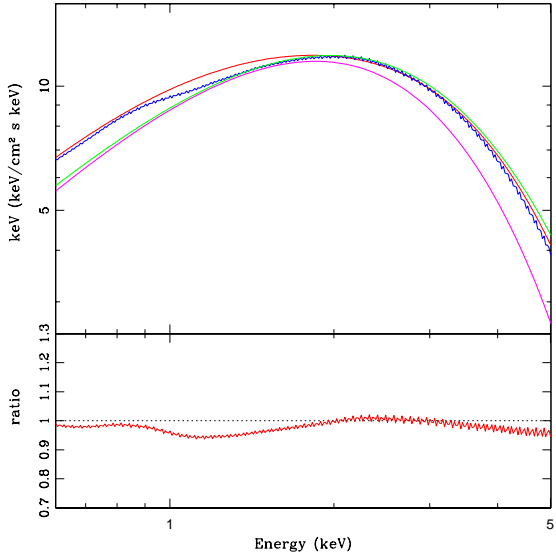}
\end{center}
\caption{Comparison of different disc continuum models with Obs 3. {\sc diskbb+comptt} is plotted in red, {\sc bhspec} in blue, {\sc kerrbb} in green and {\sc diskbb} in magenta. The bottom panel shows the ratio of {\sc bhspec} to {\sc diskbb+comptt}. The phenomenological {\sc diskbb+comptt} continuum is clearly most agreeable with the data, and hence is chosen as the disc model in the followin analysis.} 
\label{disc1201}
\end{figure}

\begin{figure}
\begin{center}
\leavevmode\epsfxsize=8cm\epsfbox{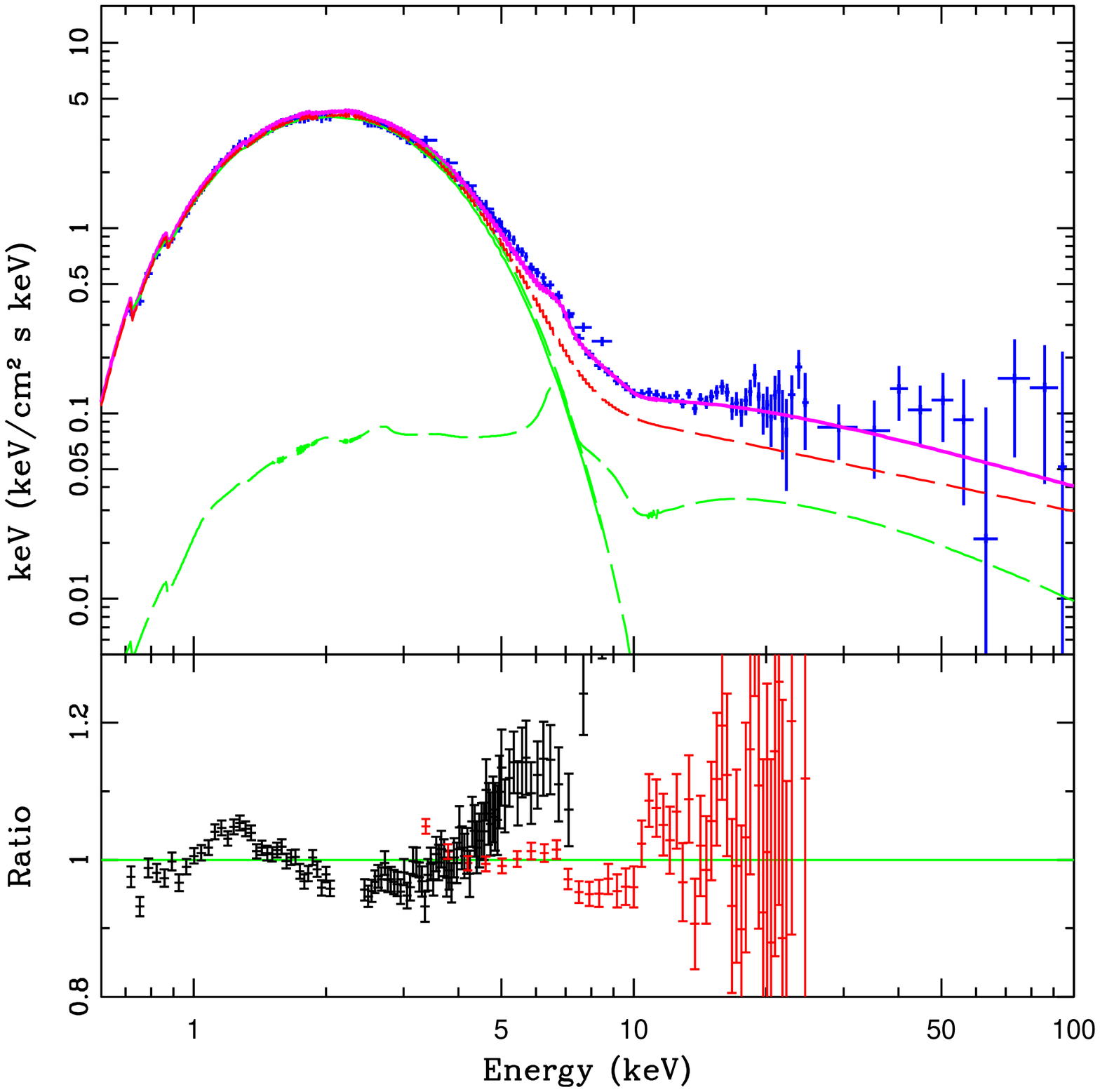}
\end{center}
\caption{The faintest disc dominated spectrum (Obs. 5) modelled with Model 3. The top panel shows the data+model, along with the model components (disc+reflection). The model residuals are plotted in the bottom panel.} 
\label{1701_bhspec}
\end{figure}

\begin{figure}
\begin{center}
\leavevmode\epsfxsize=8cm\epsfbox{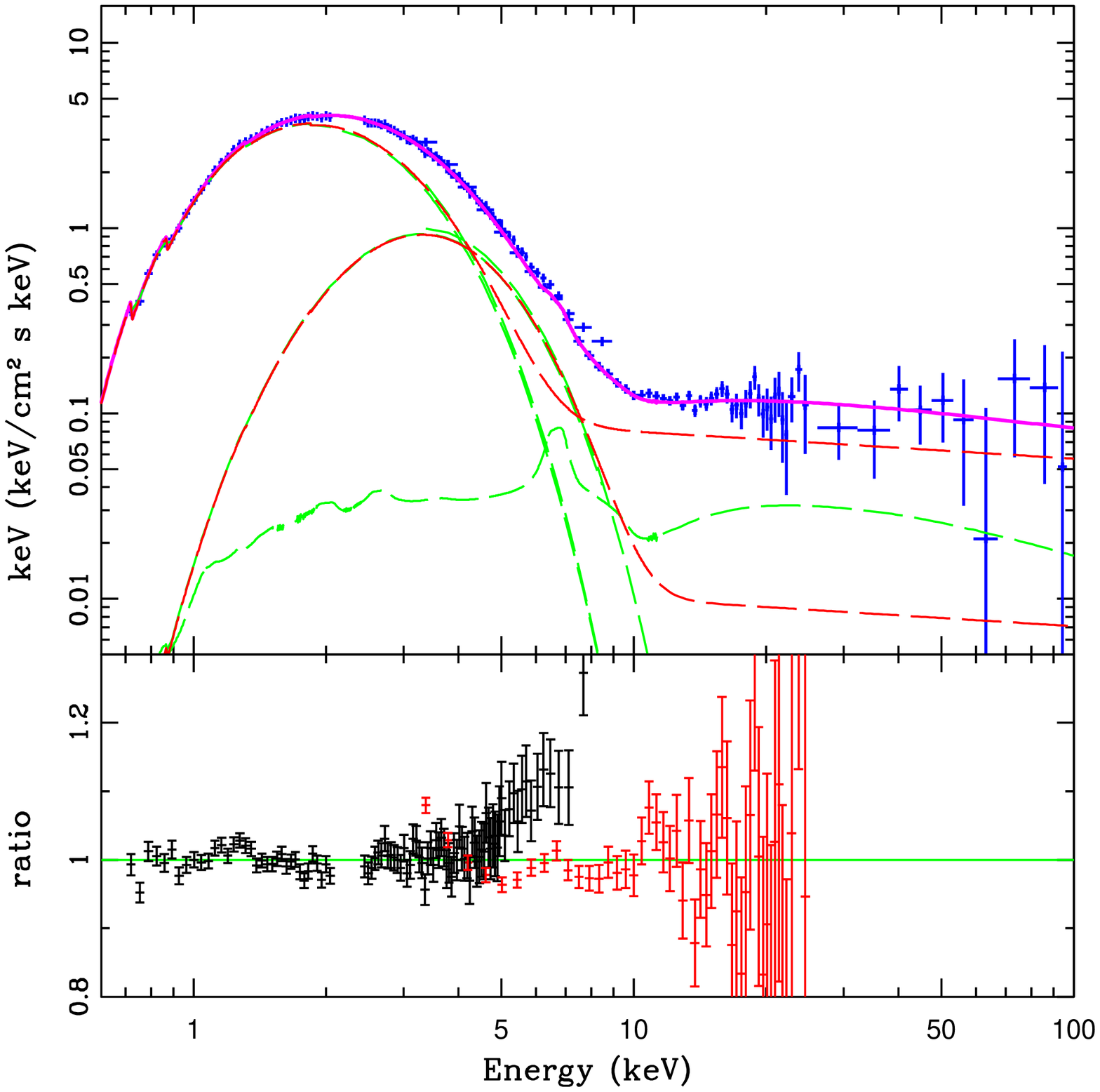}
\end{center}
\caption{ Same as Fig ~\ref{1701_bhspec}, now modelled with Model 4. } 
\label{1701_conv}
\end{figure}

Figures~\ref{1201_bhspec} -- \ref{1701_conv} show the deconvolved
spectra for both model fits to observations 1 and 5. 
Unlike the more luminous spectrum discussed in the
previous section, there is now a clear mismatch above 4~keV between
the pn and PCA spectra. This is most likely due to the lack of
background subtraction in the burst mode data, which is now beginning
to become an issue at high energies for this lower luminosity
spectrum. We discuss this in more detail in Appendix~\ref{bkg}. 

The remaining disc dominated spectrum, Obs. 1, is very similar to the 
Obs. 3 so we do not show the
deconvolved spectra, but give details of the fits in Tables~\ref{model3} and \ref{model4}. 
Again, the main difference between Model 3 and 5 is that the
{\sc bhspec} models predict smeared absorption features which are not
seen in the best fit {\sc diskbb+comptt} models.

\begin{figure}
\begin{center}
\leavevmode\epsfxsize=8cm\epsfbox{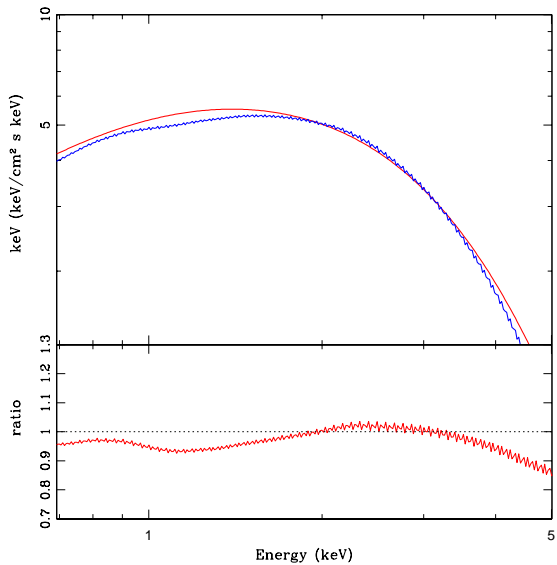}
\end{center}
\caption{Same as Fig~\ref{disc1201} for Obs. 5, now showing the phenomenological model {\sc diskbb+comptt} in red and the best theoretical model {\sc bhspec} in blue. The bottom panel shows the ratio of {\sc bhspec} to {\sc diskbb+comptt}. } 
\label{disc1701}
\end{figure}

\section{The soft intermediate state spectra (SIMS)}

A significant advantage to using the {\small {\sc simpl}} model is
that it also allows us to fit spectra with strong tails in addition to
the thermal dominated ones. We now use the same models (3 and 4) to
go through the SIMS spectra in order of the strength of the tail.

\begin{figure}
\begin{center}
\leavevmode\epsfxsize=8cm\epsfbox{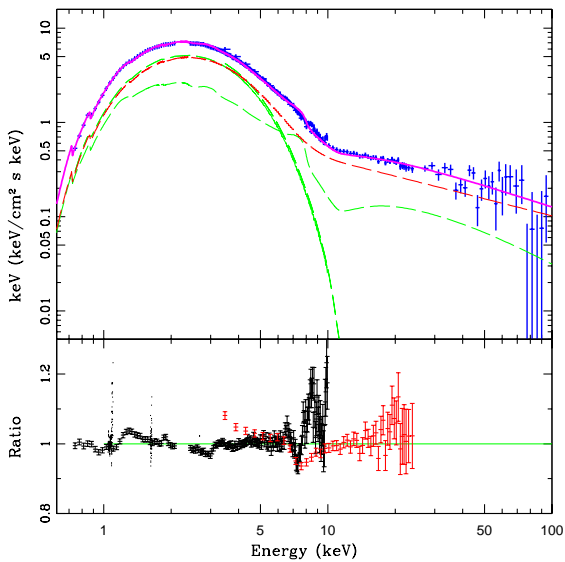}
\end{center}
\caption{The 2002/2003 SIMS spectrum (Obs. 2) modelled with Model 3. The top panel shows the data+model, along with the model components. The model residuals are plotted in the bottom panel.} 
\label{miller_bhspec}
\end{figure}

\begin{figure}
\begin{center}
\leavevmode\epsfxsize=8cm\epsfbox{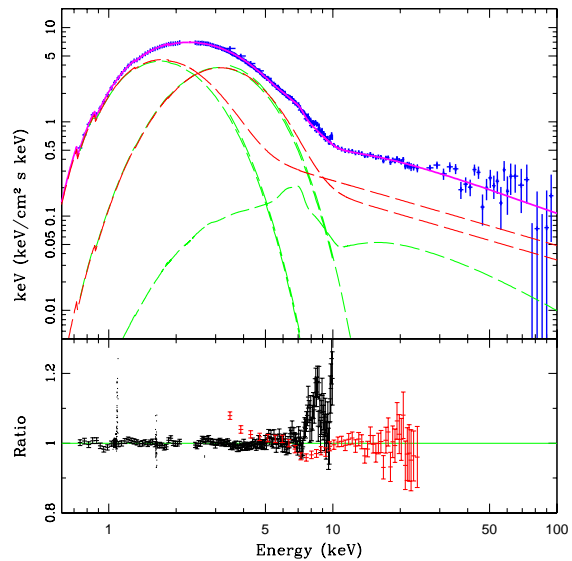}
\end{center}
\caption{Same as Fig~\ref{miller_bhspec}, modelled with Model 4.} 
\label{miller_conv}
\end{figure}

\subsection{The 2002/2003 outburst: Obs. 2}

The tail carries roughly 25\% of the total luminosity, and Figure~\ref{l-t}
shows that it is slightly steeper than the tail seen in the disc
dominated spectra.  Our best-fit for Model 3 gives a very poor
$\chi^2$ of 798/199 d.o.f., while Model 4 gives a better fit with
$\chi^2_\nu$=451/196. These are shown in Figures~\ref{miller_bhspec}
and ~\ref{miller_conv}. Both have clear residuals above 6~keV in the
pn but not the PCA. The residuals in Model 3 look like there may be an 
absorption line at $\sim 7$~keV, but this feature disappears with the
different continuum of Model 4 and instead there is a line-like residual
at $\sim 8$~keV. This could be due to the copper line in the pn, but
no such line is seen in the residual background in Fig~\ref{background}. However, 
even Model 4 struggles to fit the shape of the spectrum well, 
though some of this could be that the disc ionisation
parameter hits its upper limit (log$\xi=4.00_{-0.04}$), i.e. the model
does not extend to high enough ionisation to fit the data. 

These data were previously analysed (though with older versions of the
SAS which did not include CTI corrections) by e.g. Miller et
al. (2004;2006;2008;2009) and Reis et al. (2008). These studies
observed a very skewed iron line at $\sim$6.4~keV, indicating an
almost maximally spinning black hole. This is in sharp contrast to the
results from either Model 3 or 4 in our spectral fits, where the
derived inner radius is quite large at $R_{in}= 
34^{+10}_{-17}R_{g}$. We return to this point in 
Section~\ref{discussion}.

\subsection{The 2007 outburst: Obs. 4}

\begin{figure}
\begin{center}
\leavevmode\epsfxsize=8cm\epsfbox{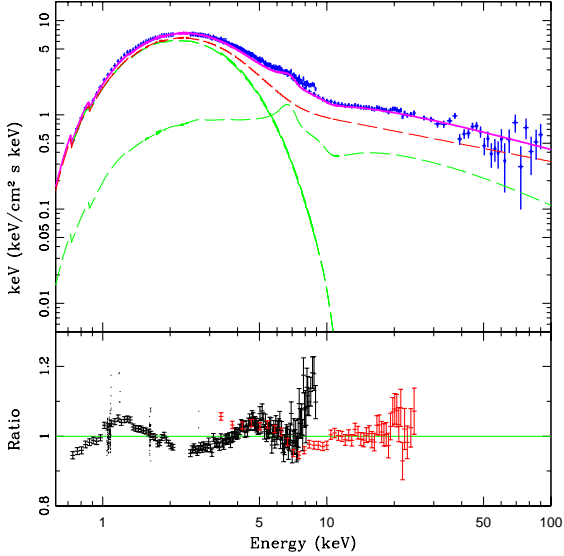}
\end{center}
\caption{The 2007 SIMS spectrum (Obs 4.) modelled with Model 3. The top panel shows the data+model, along with the model components. The model residuals are plotted in the bottom panel.} 
\label{1301_bhspec}
\end{figure}

\begin{figure}
\begin{center}
\leavevmode\epsfxsize=8cm\epsfbox{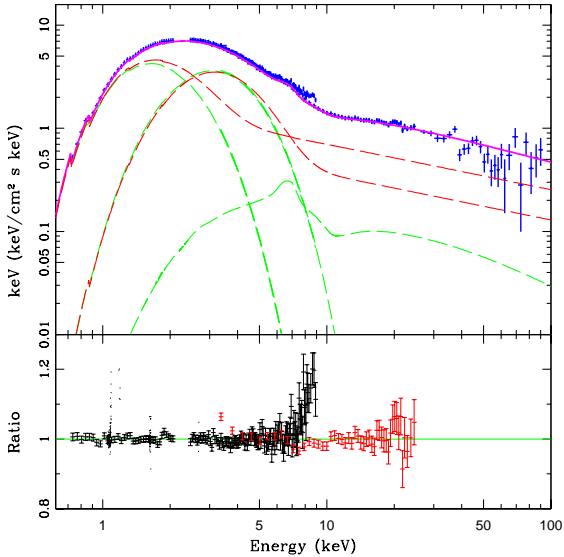}
\end{center}
\caption{Same as Fig~\ref{1301_bhspec}, modelled with Model 4. } 
\label{1301_conv}
\end{figure}

Figure~\ref{all_spectra} shows that the strongest tail is in the 
2007 outburst, in observation 4 (green). 
Results from fits with Models 3 and 4 are again shown in Tables~\ref{model3} and
\ref{model4}, and the spectra shown in Figures~\ref{1301_bhspec} and
~\ref{1301_conv}. 

Model 3 gives an unacceptable $\chi^2$=968/186 d.o.f. The predicted
continuum falls substantially below the HEXTE points and the derived
column is lower than for the other datasets, at $N_H=5.0^{+0.04}_{-0.02}\times 10^{21}$~cm$^{-2}$. Model 4, on the other hand, gives a
best-fit $\chi^2$=297/183 d.o.f (though there is still a mismatch in residuals between the pn and the PCA above 7~keV), with hydrogen column of
$N_H=5.45\times 10^{21}$~cm$^{-2}$, more consistent with the other
datasets. The reflected continuum has moderate solid angle
of $\Omega/2\pi=0.63^{+0.16}_{-0.10}$ and again, 
the smearing is not at
all extreme, with $R_{in}=42^{+18}_{-30} R_{g}$.
This reflection is highly ionised, but is within the range of
tabulated models at log~$\xi=3.9\pm 0.1$. Since these data are 
brighter, it seems unlikely that the ionisation parameter is truly 
higher in the previous soft intermediate state above (Obs. 2). Instead, 
this probably indicates that the previous data are more complex than
the model fit. 

\section{Black hole spin}
\label{discussion}

\subsection{Disc continuum fits: {\sc bhspec}}

The {\sc bhspec} model fits directly for black hole spin from the
dominant disc component for a given mass, distance and inclination.
These system parameters are quite poorly known for GX~339$-$4
(see e.g. Kolehmainen \& Done 2010), but fixing these
at reasonable values of $10M_\odot$, $8$~kpc and $60^\circ$,
respectively, gives derived spin values which are low to moderate
($a_*=0.1-0.5$). This is as expected from the previous fits to the
\textit{RXTE} PCA data (Kolehmainen \& Done 2010). However, the derived spin is
not constant, even when restricted to the disc dominated spectra
(Table~\ref{model3}) due to the mismatch between the broad absorption
features from oxygen and iron L predicted by
{\sc bhspec} and the data (Figs~\ref{disc1201} and \ref{miller_bhspec}). The excellent statistics of
the \textit{XMM-Newton} data means that these 5--10\% residuals drive
the Comptonised continuum to a much steeper index to compensate for
this lack of flux at low energies. This level of mismatch between the
data and model is enough to distort the entire fit.

These fits (Table~\ref{model3}) also give an estimate for black hole spin from
the iron line profiles. However, since the fits are so distorted, these
are clearly not reliable.

\begin{figure*}
\begin{tabular}{cl}
\leavevmode\epsfxsize=8cm\epsfbox{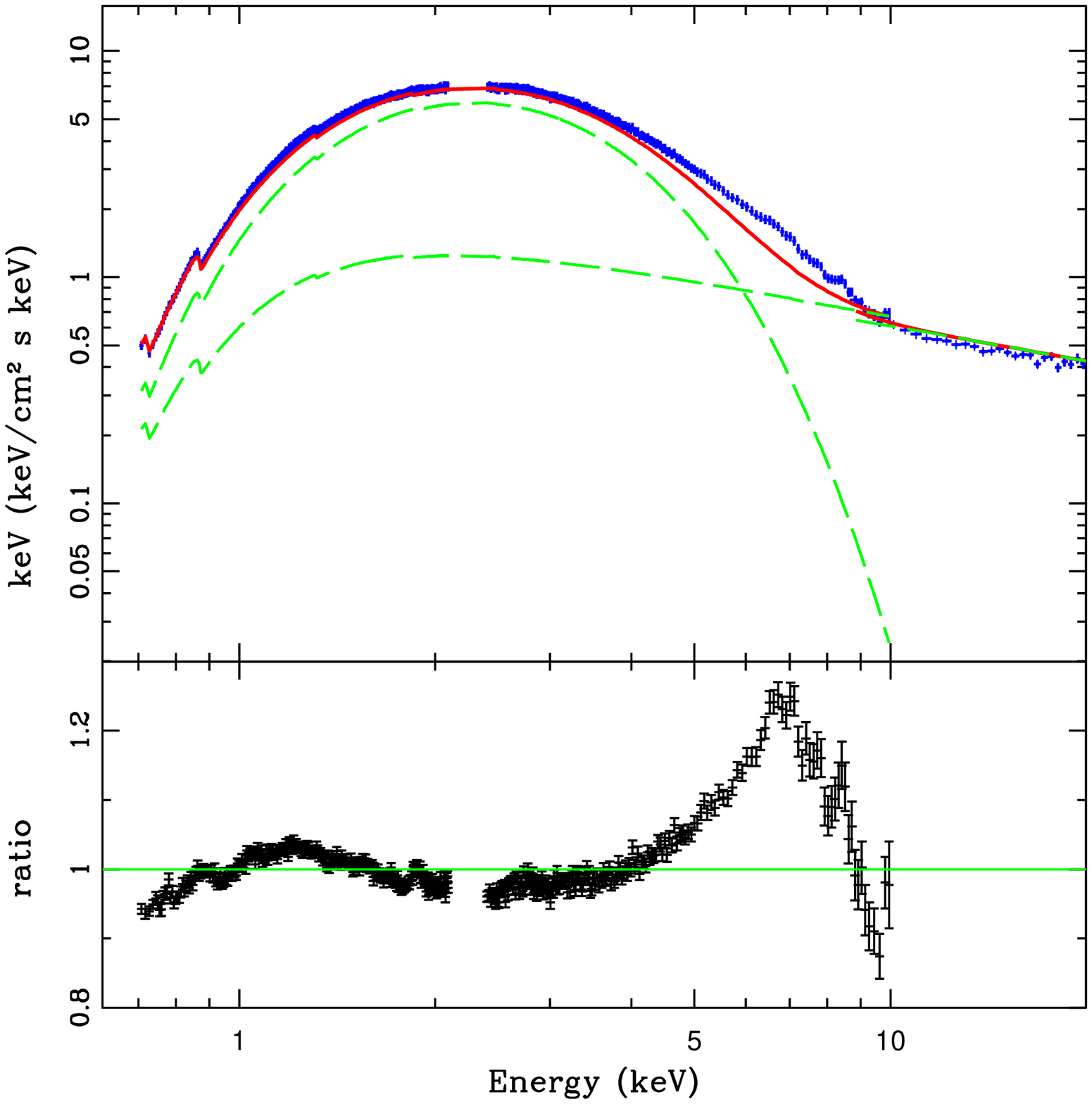} 
\leavevmode\epsfxsize=8cm\epsfbox{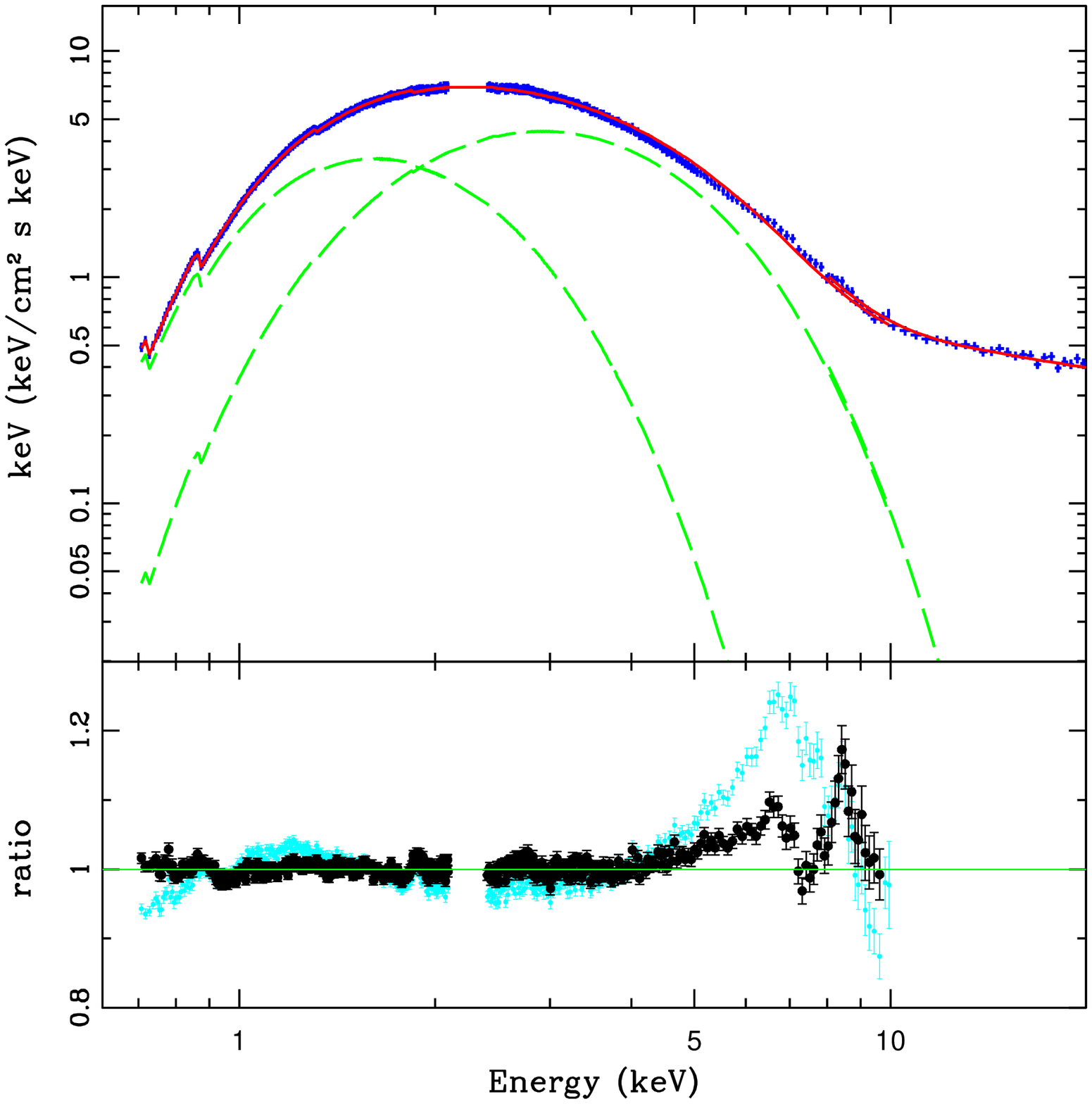} 
\end{tabular}
\caption{The 2002/2003 SIMS observation (Obs. 2) fitted with different set of models to illustrate the dependency of emission line residuals to the chosen continuum model. The continuum is fitted by excluding 4--7~keV. Only EPIC pn residuals are plotted for clarity. \textit{a) Left panel}: {\sc diskbb$+$powerlaw}. The residuals show a broad emission line feature at $\sim$6.5~keV. \textit{b) Right panel}: The same spectrum modelled with Model 4. Residuals show narrow line features (black) and as a comparison the residuals from the \textit{left panel} (cyan).} 
\label{ratio}
\end{figure*}

\subsection{Iron line}

The disc continuum is much better modelled using the phenomenological
{\sc diskbb+comptt} description. With this, the amount of smearing of the
reflected emission should be indicative of the inner radius of the disc.
However, none of our spectral fits give a line which is so broad as to require that
the disc extends down to the last stable orbit around even a 
Schwarzschild black hole. This is in sharp contrast to the claim 
of Miller et al. 2004; Reis et al. 2008 for the less extreme
soft intermediate state spectrum (Obs. 2), that 
the iron line is so broad as to require 
that the black hole in GX~339$-$4 has high spin, of $a_*$=0.935.
This value is also in conflict with the upper limit for the spin 
of $a_*\leq0.9$ in GX~339$-$4 (Kolehmainen \&
Done, 2010). 
We repeat their analysis on this specific dataset (Obs. 2, combined with \textit{RXTE} PCA data), and fit the spectrum, excluding the 4--7~keV (i.e. Fe line) region,
with a simple diskbb$+$power law. We then plot residuals including
the Fe line bandpass.  Our residuals indeed show a very similar
line profile to their results, with an extremely broad red wing
to the iron line (Figure ~\ref{ratio}a). Furthermore, we also add a relativistic emission line model ({\sc laor}; Laor 1991) to describe the Fe line more physically. Again we see similar results in terms of a very small disc inner radius, low inclination and a very centrally peaked emissivity.

However, 
modelling complex spectral curvature such as this is a delicate
task. Fig~\ref{all_spectra} shows that the iron line is in an area where the curvature
changes from being dominated by the disc to being dominated by the tail.
The 'iron line' residual will then be strongly affected by a small change
in the continuum model. We show this explicitly by fitting our best
continuum model ({\sc diskbb+comptt} for the disc, convolved 
with {\sc simpl} for the tail) to the data, again excluding the Fe line region. 
Figure ~\ref{ratio}b then shows the same as Fig~\ref{ratio}a but with the
continuum modelled using Model 4. We overplot the previous residuals
for direct comparison and change in shape of the derived line 
profile is immediately apparent.

The line profile derived by this method is clearly very sensitive to a
change in underlying continuum shape. The analysis of the disc
dominated spectra in Section~\ref{discy} shows that the disc continuum is
substantially broader than a {\sc diskbb} due mainly to 
relativistic smearing of the disc continuum and radiative transfer through
the disc photosphere. Fitting a {\sc diskbb} continuum then forces
a broad residual into the data from the poor match to the disc
continuum. However, this broad residual on the tail of the disc spectrum
is in the same energy band as the expected red wing of the iron line. 
We caution that the derived iron line shape depends on the assumed
continuum, and that this is especially important where the continuum
curvature changes rapidly under the iron line. This is the case for
all the bright disc dominated and soft intermediate spectra considered
here. 

Table~\ref{model4} shows that with our best fit model the line is not extremely 
smeared in these data (consistent with Fig~\ref{ratio}b), 
with  $R_{in}= 34^{+10}_{-17}R_{g}$. 
This does not mean that we infer the disc to be
truncated. Quite the contrary, the unfolded spectrum has a clear disc
shape below $\sim$6~keV and the disc contributes to $\sim$75\% of the
total luminosity. Our fit also requires that the reflecting material is 
highly ionised, as expected from the high effective temperature of the
disc. It seems more likely that the inner disc is so highly ionised that 
iron is completely stripped, so no longer contributes to the atomic 
features which are the best tracer of the reflecting material. 
We also caution that while this model is our best fit, our iron line
parameters will also depend on our assumed continuum form. 
While our reflection is derived from the best currently available 
models (Ross \& Fabian 2005), these are calculated for the lower
density and temperatures of AGN discs. Newer reflection models
including the effect of collisional ionisation should instead be used
(Ross \& Fabian 2007), but
we caution that these alone will not solve the sensitivity of the iron line
profile to the underlying continuum shape.

\begin{table*}
\begin{tabular}{l|l|l|l|l|l|l|l|l|c}
 \hline
 \multicolumn{7}{|c|} {\sc disc:bhspec} {\sc Comptonisation + reflection}  \\	
 \hline
  Obsid & $N_H$($\times 10^{21}$)& $L/L_{\rm{Edd}}$  & $a_{*}$ & $\Gamma$ & $R_{in}$ ($R_{g}$) & log$\xi$ & $f=\Omega/2\pi$ & $\chi^{2}$/ d.o.f
 \\
 \hline
 \hline
0093562701  & $5.31^{+0.01}_{-0.01}$ &$0.26\pm0.01$  & $0.20\pm0.003$ & $3.21^{+0.01}_{-0.03}$& $26^{+7}_{-6}$ & $3.30^{+0.01}_{-0.04}$ & $5.70^{+0.65}_{-0.15}$  & 721/198\\

0156760101  & $5.58^{+0.10}_{-0.05}$ &$0.14\pm0.01$  & $0.51\pm0.01$ & $2.80^{+0.02}_{-0.02}$& $16^{+2}_{-2}$ & $3.36^{+0.02}_{-0.02}$ & $0.66^{+0.04}_{-0.04}$  & 798/199\\

0410581201  & $5.50^{+0.03}_{-0.05}$ &$0.15\pm0.01$  & $0.54\pm0.01$ & $3.19^{+0.09}_{-0.05}$& $33^{+31}_{-9}$ & $3.09^{+0.01}_{-0.05}$ & $6.94^{+0.42}_{-0.55}$  & 426/186 \\

0410581301  & $5.00^{+0.04}_{-0.02}$ &$0.17\pm0.01$  & $0.18\pm0.01$ & $2.51^{+0.01}_{-0.05}$& $32^{+12}_{-7}$ & $4.00_{-0.31}$ & $0.72^{+0.04}_{-0.09}$  & 968/186 \\

0410581701  & $5.25^{+0.03}_{-0.03}$ &$0.12\pm0.01$  & $0.21^{+0.02}_{-0.01}$ & $2.35^{+0.01}_{-0.04}$& $33^{+11}_{-8}$ & $3.31^{+0.02}_{-0.01}$ & $3.34^{+3.60}_{-2.89}$  & 519/178 \\
\hline
\end{tabular}
\caption{The best-fit parameter values of Model 3. The errors quoted are shown for illustrative purposes only, in summary of the freedom of fit.}
\label{model3}
\end{table*}

\begin{table*} 
\begin{tabular}{lc|l|l|l|l|l|l|l|l|l|c} 
  \hline
\multicolumn{8}{|c|} {\sc disc:diskbb+comptt} {\sc Comptonisation + reflection}  \\
  \hline
 Obsid & $N_H$($\times 10^{21}$) & $T_{in}$ (keV) &$kT_{e}$  & $\tau$ & $\Gamma$  & $R_{in}$ ($R_{g}$) & log$\xi$ & $f=\Omega/2\pi$ & $\chi^{2}$/ d.o.f
 \\
 \hline
 \hline
0093562701  & $5.30^{+0.01}_{-0.03}$  & $0.57^{+0.01}_{-0.002}$ & $0.76^{+0.002}_{-0.003}$ & $60^{+5}_{-24}$ & $2.45^{+0.06}_{-0.02}$ & $101^{+298}_{-37}$ & $3.88^{+0.10}_{-0.11}$ & $0.77^{+0.04}_{-0.11}$  & 276/195  \\

0156760101  &  $5.35^{+0.01}_{-0.02}$  & $0.49^{+0.01}_{-0.002}$ & $0.74\pm0.01$ & $33^{+1}_{-3}$ & $2.69^{+0.03}_{-0.01}$ & $34^{+10}_{-17}$ & $4.00_{-0.04}$ & $0.76^{+0.04}_{-0.03}$  & 451/196 \\

0410581201  &  $5.25^{+0.03}_{-0.04}$  & $0.59\pm0.01$ & $0.77\pm0.01$ & $120^{+80}_{-50}$ & $2.41^{+0.17}_{-0.10}$ & $64^{+99}_{-30}$ & $3.13^{+0.30}_{-0.18}$ & $0.46^{+0.10}_{-0.13}$  & 210/183 \\

0410581301  &  $5.45^{+0.02}_{-0.07}$  & $0.46\pm0.01$ & $0.71\pm0.01$ & $30^{+3}_{-3}$ & $2.44^{+0.02}_{-0.03}$ & $42^{+18}_{-30}$ & $3.90^{+0.10}_{-0.10}$ & $0.63^{+0.16}_{-0.10}$  & 297/183\\

0410581701  &  $4.87^{+0.04}_{-0.05}$  & $0.58\pm0.01$ & $0.73\pm0.01$ & $65^{+117}_{-65}$ & $2.15^{+0.18}_{-0.05}$ & $235^{+164}_{-136}$ & $3.09^{+0.17}_{-0.12}$ & $0.62^{+0.10}_{-0.17}$  & 323/175 \\

\hline
\end{tabular}
\caption{The best-fit parameter values of Model 4. The errors quoted are shown for illustrative purposes only, in summary of the freedom of fit.}
\label{model4}   
\end{table*}

\section{Discussion and conclusions}

The new calibration of the XMM-Newton EPIC pn burst mode allows
detailed spectral fitting of bright black hole binaries. We use these 
data together with (mostly) simultaneous RXTE data to 
explore the shape of the disc dominated and soft intermediate state
0.7--200~keV spectra from GX~339$-$4. This is an important object 
to understand as there are conflicting measures of the black hole
spin in this object, with disc continuum fitting of disc dominated states
from RXTE showing an upper limit of $a_*<0.9$ (Kolehmainen \& Done 2010) 
while the iron line profile in an XMM-Newton 
dataset from a soft intermediate state gives $a_*=0.935$ (Miller et al. 2004;
Reis et al. 2008). 

We find that while the disc dominated states are well fit with 
the simple {\sc diskbb} model (together with a tail to high energies and its
reflection) in the PCA, the lower energy extent of the CCD bandpass
shows that this is not a good representation of the disc emission.
This is as expected, as the disc continuum should also be smeared
by relativistic effects, making it substantially broader. We fit the
best current disc models to the data instead. These include 
full radiative transfer through the disc photosphere as well as relativistic
smearing ({\sc bhspec}: Davis et al. 2005). However, these
are not a good match to the data either, as although they are
broader, they predict smeared
atomic absorption features at oxygen/iron L which are not present 
in the data. This mismatch is also seen in \textit{Suzaku} data from LMC X-3
(Kubota et al. 2010), making it unlikely to be a residual calibration 
feature of \textit{XMM-Newton}. Instead, it could be due to irradiation of
the photosphere by the weak high energy tail. 

Whatever its origin, the disc dominated spectra clearly show that the 
observed disc spectrum can only be currently well matched by 
phenomenological models. We use the same phenomenological description 
to fit the disc component in the soft intermediate state. This broader
disc spectrum has a profound effect on the derived iron line profile, as the
line energy is close to where disc and tail components have equal
flux. A small change in the disc spectrum then makes a large change
in the residual flux at the iron line energy. We find that we cannot 
constrain black hole spin in these data with our continuum model
as the line profile (which is already smeared by Compton scattering
in the highly ionised reflector) is not strongly relativistically smeared,
so arises predominantly from radii which are rather larger than 
even the last stable orbit around a Schwarszchild black hole. This
probably indicates that the inner disc is so highly ionised that iron is
completely stripped and hence does not produce any characteristic
atomic features. 

We caution that using simple continuum models for complex spectra
will result in broad residuals in the data simply from deficiencies in the
continuum modelling rather than giving a robust, model independent
way to see the iron line profile. All bright black hole binary spectra are
complex, as the disc temperature is hot enough that the iron line
region is on the cross-over between the disc and tail. Only the very 
dimmest disc dominated spectra, where the disc peaks below 
$\sim 1$~keV, and the low/hard state 
have an iron line region where the continuum is mostly dominated by 
the tail. However, we also caution that the tail can also have complex
(though more subtle) curvature which may also affect the line
residuals. 

\section{Acknowledgements}

We thank Matteo Guainazzi for the very useful discussions on calibration.
MK also acknowledges the support of the Finnish Cultural Foundation.

\appendix

\section{Burst mode background}
\label{bkg}

\begin{figure*}
\begin{center}
\leavevmode\epsfxsize=16cm\epsfbox{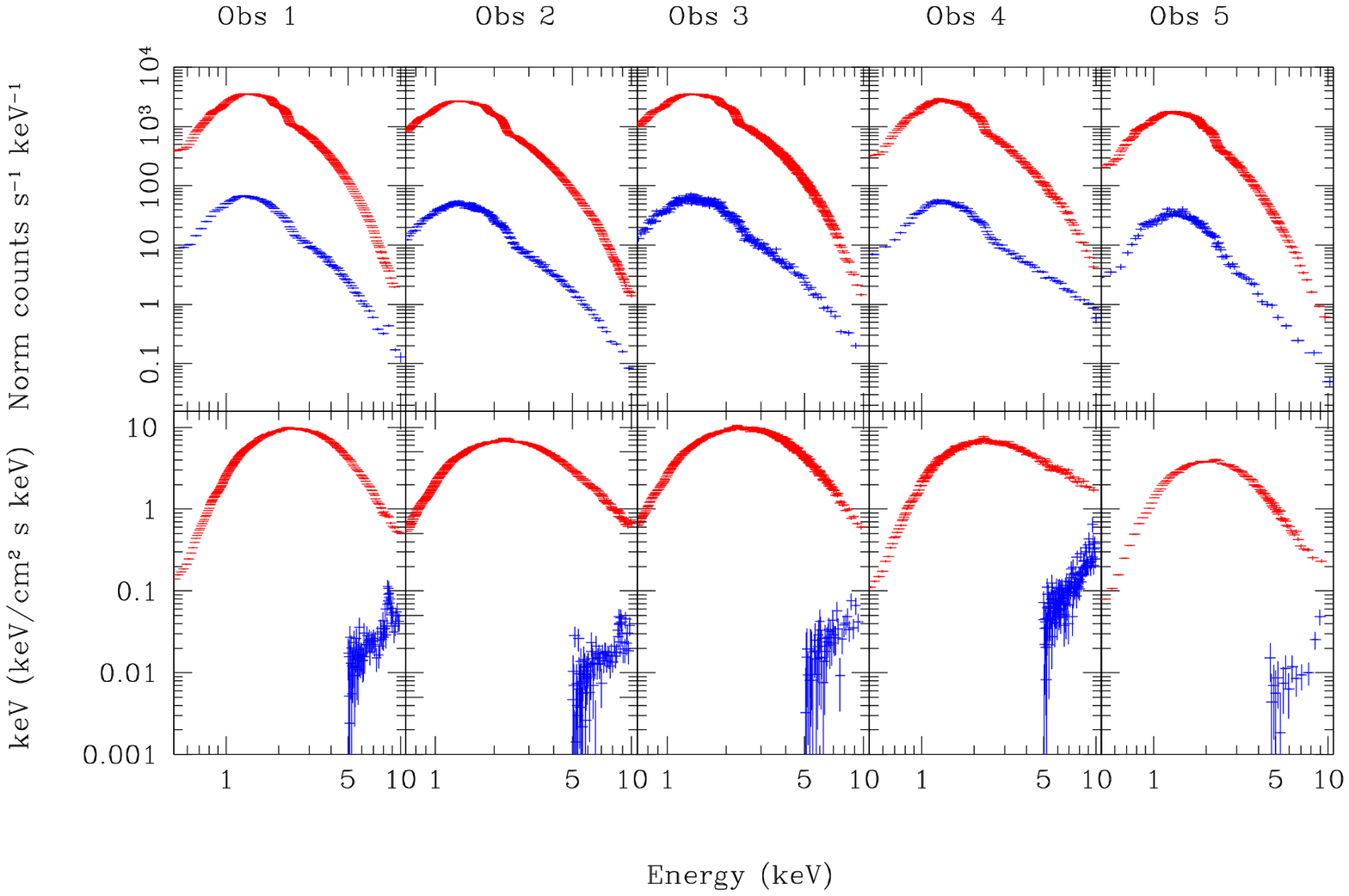}
\end{center}
\caption{Background analysis for the EPIC pn burst mode. \textit{Top panel}: The source spectrum (in red), plotted with the extracted background (blue) in detector counts. The shape of the background plainly follows the shape of the source, indicating that much of the 'background' is in fact contaminated by the source. \textit{Bottom panel}: The background spectrum (in blue) corrected with the source spectrum. Observations 1 and 2 show the clearest copper lines at $\sim$8~keV, while the 'background' level in observation 4 is clearly strongest of the sample. } 
\label{background}
\end{figure*}

Background subtraction of bright sources in burst (and timing) modes is 
very challenging. The bright source contaminates the whole CCD chip, 
and hence background was not subtracted in the reduction phase of this 
analysis. 
To illustrate this effect, we extract a 'background' spectrum for all of our observations. The extraction region was selected 
as a strip of RAWX [3:10] and RAWY [1:160], near the edge of the chip, 
away from the source itself. 

Figure~\ref{background} (top panel) shows the source spectrum (red), 
plotted with the background (blue) in detector counts. The shape of the background clearly 
mimics that of the source, indicating that much of the extracted 
'background' is in fact contaminated by the source (see also Done
\& Diaz Trigo 2010 for similar issues in EPIC pn timing mode, 
and Guainazzi et al. 2010a). We attempt to constrain the true
background by subtracting a scaled version of the source spectrum from the
extracted background. We set the scale by assuming that the true
background is negligible in the 1--2~keV bandpass. The remaining background is still a slight overestimate of the
true background as high energy
photons are preferentially scattered into the wings of the point spread
function, so the spectrum at large off-axis angles will be slightly harder
than the on-axis source spectrum. We plot the resulting 
background estimate as unfolded ($\nu F_\nu$) spectra in the lower panel
of Fig.~\ref{background}. 

The background above 5~keV from singles in the pn full frame mode 
is typically at a level of 0.1 counts/s/keV, with a strong copper line 
superimposed (Freyberg et al. 2006: XMM-SOC-CAL-TN-0068\footnotemark[1]). Adding the doubles (burst mode loses one spatial dimension,
so cannot distinguish between singles and doubles in the compressed 
direction) would increase this by a factor $\sim 2$. This is roughly the 
level seen at 10~keV in all the spectra  (upper panel) 
except for the one with the  strongest tail (Obs. 4). The lower panel
shows the source subtracted $\nu F_\nu$ looks similar to that 
expected from the background, with a strong copper line at 8~keV
in all the disc dominated spectra. However, the 'background' level is
progressively larger, swamping the line, in the spectra with stronger
tails. This probably indicates that there is residual contamination from the 
source, and this contamination gets stronger for stronger hard spectra. 

Thus the residual background in the faintest disc dominated spectrum is
probably closest to the 'true' background for singles plus doubles in 
burst mode. This is 20\% of the source flux at 8--10~keV. The bright
disc dominated spectra are a factor $\sim$ 3--4 brighter at 10~keV 
(see Fig~\ref{all_spectra}), so a similar background here would give a 5\% excess 
at high energies in the pn. The residuals to the best fit model 
appear to be slightly larger than this (Fig~\ref{1201_conv}) which may indicate 
remaining cross-calibration uncertainties between the pn and PCA 
(Guainazzi et al. 2010, Weisskopf et al. 2010).  

The clear conclusion is that the background in burst mode is not 
necessarily negligible, especially for very soft spectra. There is 
no way to estimate this reliably from the data, so 
offset pointings are required. 

\label{lastpage}

\end{document}